\documentclass[groupaddress,prb,twocolumn]{revtex4}

\usepackage{hyperref}
\usepackage{amsmath}
\usepackage{amssymb}
\usepackage{graphicx}
\usepackage{verbatim}

\begin{document}

\title{Non-local effects in effective medium response of nano-layered meta-materials}

\author{Justin Elser}
\author{Viktor A. Podolskiy}

\affiliation{Department of Physics, Oregon State University, \\ Corvallis, Oregon 97331}

\author{Ildar Salakhutdinov}
\author{Ivan Avrutsky}

\affiliation{Department of Electrical and Computer Engineering, Wayne State University, \\  Detroit, Michigan 48202}


\begin{abstract}
We analyze electromagnetic modes in multi-layered nano-composites and demonstrate that the response of a majority of realistic layered structures is strongly affected by the non-local effects originating from strong field oscillations across the system, and is not described by conventional effective-medium theories. We develop the analytical description of the relevant phenomena and confirm our results with numerical solutions of Maxwell equations. Finally, we use the developed formalism to demonstrate that multi-layered plasmonic nanostructures support high-index volume modes, confined to deep subwavelength areas, opening a wide class of applications in nanoscale light management. 
\end{abstract} 

\maketitle

\noindent Nanolayered composites have been recently proposed to serve as negative index systems, super- and hyper-lenses, photonic funnels, and other nanophotonic structures\cite{superlens,hyperlens,shvetsPRB,fanNIM,engheta,anisotropy,funnels,Antos2006,Verney2004,Korob2006,Bennin1999}. The typical thickness of an individual layer in these ``artificial'' (meta-) materials is of the order of $10 nm$. Since this size is much smaller than optical (or IR) wavelength, it is commonly assumed that the properties of the multilayered composites are well-described by the effective medium theory (EMT)\cite{landauECM,SmithMeta}. In this Letter, we analyze the modes of realistic multi-layered structures and show that the conventional EMT fails to adequately describe these modes due to the metamaterial analog of spatial dispersion -- strong variation of the field on the scale of a single layer. We derive a non-local correction to EMT, bridging the gap between metamaterial- and photonic crystal-regimes of multi-layered media, and use numerical solutions of Maxwell equations to verify our results. Finally, we use the developed technique to identify {\it volume} metamaterial modes confined to nanoscale areas. 

While the formalism developed below is applicable to the composites with arbitrary values of permittivities operating at different frequency ranges (UV, visible, IR, THz), here we illustrate our approach on the optical response of a two-component plasmonic nanolayered composite, which has been suggested for a variety of future beam-steering and imaging systems\cite{hyperlens,fanNIM,anisotropy}. The schematic geometry of such a structure, containing alternating layers of materials with permittivities, $\epsilon_1$, $\epsilon_2$ and (average) thicknesses $a_1$ and $a_2$ respectively, is shown in Fig.\ref{system}. In the analytical results presented below, we mostly focus on the propagation of TM waves, which are responsible for plasmon-assisted phenomena; and only briefly discuss the implications for TE modes. It is straightforward to generalize the presented technique for mixed waves as well as for multi-component structures. In the selected geometry, $ x$ coordinate axis is perpendicular to the plane defined by layer interfaces, while $y$ and $ z$ axes are parallel to this plane; the direction of $z$ axis is chosen so that that the electromagnetic waves propagate in $x,z$ plane. 

\begin{figure}[t]
\centering
\includegraphics[width=4cm]{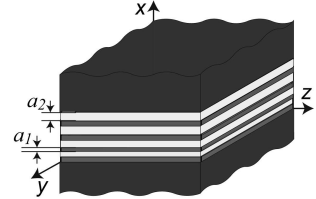}
\caption{Schematic geometry of a planar nanolayer-based meta-material, surrounded by two cladding layers}
\label{system}
\end{figure}

The majority of realistic designs of layered nanoplasmonic structures\cite{fanNIM,anisotropy,hyperlens,enghetaJOSAB} rely on the {\it metamaterial regime}, when the typical layer thickness is much smaller than the free-space wavelength $\lambda$ so that surface plasmon polaritons propagating on different metal-dielectric interfaces are strongly coupled to each other. The optical properties of the metamaterial structure in this strong coupling regime could be related to some effective permittivities. The existence of these effective parameters is important from both fundamental and applied standpoints. Thus, effective-medium description provides one with an insight into the physics behind the optical response of the structure; it also significantly reduces the computational efforts needed to simulate the electromagnetic response of the system, simultaneously increasing the accuracy of these simulations; finally, it can be easily utilized to provide an important link between the properties of (easily fabricatable) systems with a few layers, with the ones of (more practical) macroscopic multi-layered structures.

Apart from the wavelength, two more independent length-scales can be identified in the system -- the one of the typical layer thickness $a\sim a_1,a_2$, and the one of the typical field variation $L$. Since the introduction of effective permittivity $\epsilon^{\rm eff}$ requires some kind of field averaging, independence of $L$ and $\lambda$ yields to a fundamental difference between the metamaterial\cite{SmithMeta} and ``conventional'' effective-medium\cite{landauECM} responses of nano-composites. As we show below, in nanoplasmonic layered structures $L\lesssim\lambda$ so that $\epsilon^{\rm eff}$ will have non-local corrections. 

As any optical system, the multi-layered composite can be described by the behavior of its resonant (eigen) modes. Each such mode is characterized by its {\it effective modal index}, given by $n_{\rm eff}=k_z c/\omega$ with $k_z$ and $c$ being the modal wavevector and speed of light in the vacuum respectively. An arbitrary wave propagating through the system can be represented as a linear combination of different modes. Note that when $n_{\rm eff}^2>0$ exceeds that of both cladding layers, the electromagnetic field of a mode is confined inside the layered structure, which behaves like a waveguide. 

To analyze the electromagnetism in the metamaterial, we numerically solve 3D Maxwell equations in the layered geometry using the Transfer Matrix Method. In this technique, described in details in \cite{Avrut2003}, the field in each layer is represented as a combination of two (plane) waves having the same dependence in $z$ direction and propagating in the opposite $x$ directions, followed by the construction of a {\it transfer matrix} describing the collective response of the multi-layered structure. The modes of the metamaterial are then related to the eigen values and eigenvectors of the transfer matrix. 

To understand the evolution of multilayered system between meta-material and effective-medium regimes, we used the transfer-matrix techniques to identify the modes of a $200$-$nm$-thick layered composite with perfectly conducting cladding layers -- essentially representing a waveguide with deep subwavelength crossection. This technique allows us to control the field variation in the direction perpendicular to the waveguide, and simultaneously enforce the ``metamaterial condition'' $a_{1,2}\ll\lambda=1.55\;\mu m$ for all nanolayered structures in our work. 

We generated $\sim 100$ ensembles of nanocomposites with $\epsilon_1=-100$ and $\epsilon_2\simeq 2$ ($Au/SiO_2$ composite). In each ensemble, we fixed the total thickness of the composite $h=200 nm$, total concentration of metal, and randomly varied thickness of individual metal layers. The variation in layer thickness was about $10\%$ of average thickness. The idea behind the ensemble generation is two-fold. First, we aim to understand the response of {\it realistic} multi-layered systems, where the total number of layers is relatively small; second, this approach gives us the opportunity to assess the tolerance of the composite properties with respect to fabrication defects. 

\begin{figure}[t]
\centering
\includegraphics[width=7 cm]{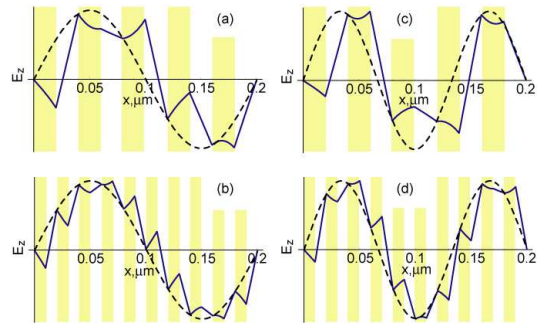}
\caption{(Color online) TMM (solid lines) and EMT (dashed lines) calculations of $E_z$ field of $\rm TM_1$(a,b), and $\rm TM_2$ (c,d) modes for metal (white)-dielectric(yellow) composites with $N_l=10$ (a,c) and $20$ (b,d) layers
}
\label{modes}
\end{figure}

The profiles of several eigen modes are shown in Fig.\ref{modes}. Note that the field across an individual layer is exponential rather than oscillatory in nature, so that high-precision arithmetic is required to find the accurate numerical solution of Maxwell equations. As the number of layers is increased (and correspondingly the thickness of an individual layer is decreased), the field distribution in the system converges to the one of the mode in a waveguide with homogeneous core. Therefore, in this regime the behavior of a multilayered composite is essentially identical to the behavior of a uniaxial anisotropic system with effective permittivity tensor $\epsilon^{\rm eff}$, given by $<D_\alpha>=\epsilon^{\rm eff}_{\alpha \beta} <E_\beta>$, with Greek indices corresponding to Cartesian components and $<>$ being the average over the multi-layer subwavelength area\cite{landauECM}. Due to axial symmetry, $\epsilon^{\rm eff}$ is diagonal, its optical axis coincides with $x$, and $\epsilon^{\rm eff}_y=\epsilon^{\rm eff}_z\equiv\epsilon^{\rm eff}_{yz}$.  

The dispersion relations of the TM and TE waves propagating in such a metamaterial\cite{anisotropy}: 
\begin{eqnarray}
\frac{\omega^2}{c^2}&=&\frac{k^2_{x}}{\epsilon^{\rm eff}_{yz}}+\frac{k^2_y+k^2_{z}}{\epsilon^{\rm eff }_{x}},
\nonumber
\\
\frac{\omega^2}{c^2}&=&\frac{k^2_{x} +k^2_y+ k^2_{z}}{\epsilon^{\rm eff }_{yz}},
\label{eqWave}
\end{eqnarray}
respectively (as noted above, $k_y=0$).

\begin{figure}[t]
\centering
\includegraphics[width=2.75in]{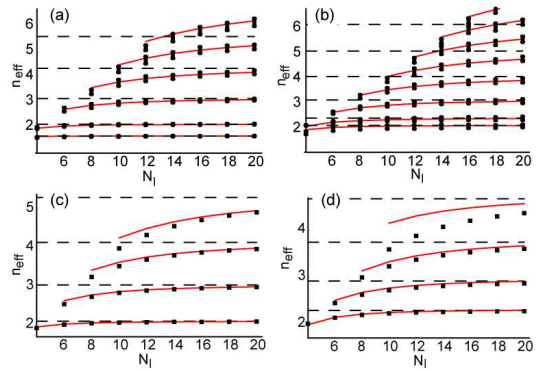}
\caption{ (Color online) Effective refractive index of waveguide modes as a function of number of layers in the core region $N_l=2h/(a_1+a_2)$, calculated using transfer matrix method (dots), ``conventional'' EMT (dashed lines), and the ``non-local EMT'' derived in this work (solid lines), for $h=200nm$-waveguide with perfectly conducting (a,b) and air (c,d) claddings; $a_1=a_2/3$ (a,c); $a_1=a_2$ (b,d); top panels correspond to $\epsilon_1=-100$; size of black bars correspond to standard deviation in $n_{\rm eff}$ as determined from our numerical simulations; bottom panels represent $\rm Au-SiO_2$ composite with $\epsilon_1=\epsilon_{\rm Au}=-114.5+11.01i$}
\label{emt_nocorr}
\end{figure}

We calculated waveguide modes for each composite in an ensemble. The results of our numerical solutions of Maxwell equations and their comparison to conventional EMT\cite{landauECM} with 
\begin{eqnarray}
\epsilon^{\rm eff}_x&=&\epsilon^{(0)}_{x} =\frac{(a_1 +a_2)\epsilon_1\epsilon_2}{a_2\epsilon_1+a_1\epsilon_2}
\nonumber
\\
\epsilon^{\rm eff}_{yz}&=&\epsilon^{(0)}_{yz} =\frac{a_1 \epsilon_{1}+a_2 \epsilon_{2}}{a_1+a_2} 
\label{eqEMT}
\end{eqnarray}
are summarized in Fig.\ref{emt_nocorr}. It is clearly seen that similar to what has been shown for fiber geometry in\cite{funnels}, the planar multi-layered composite supports highly-confined volume modes. It is also seen that while the response of all structures in a single ensemble is very alike, and therefore the introduction of effective permittivity is justified, conventional EMT fails to describe the behavior of majority of practical nanolayered composites. A reasonable agreement is present only when the number of layers $N_l=2h/(a_1+a_2)$ is very large. Note that the EMT does not work despite the fact that the condition $a_{1,2}\ll{\lambda}$ is met. 

The origin of this effect lies in a strong variation of the fields on the scale of a single layer, clearly visible in Fig.2. Similar to the strong field variation on the subatomic scale that yields non-local corrections to permittivities of homogeneous materials\cite{landauECM}, the scale separation $L<\lambda$ introduces non-locality into $\epsilon^{\rm eff}$. Note that in contrast to the case of non-local response in homogeneous structures, the microscopic (layer-specific) field in meta-material can still be described by ``local'' $\epsilon_{1,2}$; ``effective'' non-locality is present only in the effective permittivity. 

To find the non-local correction to the EMT, we start from the layered metal-dielectric structure where all metallic and all dielectric layers have the same thickness ($a_1$ may be still different from $a_2$). In this limit, the system essentially becomes a 1D photonic crystal (PC). The dispersion of the modes of this case can be related to the eigen-values problem for two-layer transfer matrix, yielding\cite{yeh} 
\begin{eqnarray}
{\cos(k_{x} [a_1+a_2])} &=\cos(k_{1} a_{1})\cos(k_{2} a_{2}) \nonumber \\
&-\gamma \sin(k_{1} a_{1})\sin(k_{2} a_{2}) 
\label{PCE}
\end{eqnarray}
where the polarization-specific parameter $\gamma$ is given by:
\begin{eqnarray}
\gamma_{\rm TM}=\frac{1}{2} \left(\frac{\epsilon_{2} k_{1}}{\epsilon_{1} k_{2}}+\frac{\epsilon_{1} k_{2}}{\epsilon_{2} k_{1}}\right), \quad \gamma_{\rm TE}=\frac{1}{2} \left(\frac{\epsilon_1}{\epsilon_2}+\frac{\epsilon_2}{\epsilon_1}\right) 
\end{eqnarray}
and
$k^2_{1,2} = \epsilon_{1,2} \frac{\omega^2}{c^2} - k^2_{z}$.

The ``conventional'' EMT regime [Eqs.(\ref{eqEMT})] can be obtained from Eq.(\ref{PCE}) through the Taylor expansion up to the second order in $|k_1 a_1|\ll 1; |k_2 a_2| \ll 1; |k_x(a_1+a_2)|\ll 1$ (see e.g.\cite{anisotropy}). Expanding the PC dispersion equation up to the next non-vanishing Taylor term yields series of modes with dispersion given by Eq.(\ref{eqWave}) and effective permittivities
\begin{eqnarray}
\epsilon^{\rm eff}_{x}=\frac{\epsilon^{(0)}_x}{1-\delta_x(k,\omega)}
\nonumber
\\
\epsilon^{\rm eff}_{yz}=\frac{\epsilon^{(0)}_{yz}}{1-\delta_{yz}(k,\omega)}
\label{eqCorr}
\end{eqnarray}
where the nonlocal corrections are given by: 
\begin{eqnarray}
\delta_x&=&\frac{a_1^2 a_2^2 (\epsilon_1-\epsilon_2)^2 {\epsilon^{(0)}_{x}}^2 }{12 (a_1+a_2)^2 \epsilon_1^2 \epsilon_2^2}  \left( \epsilon^{(0)}_{yz}\frac{\omega^2}{c^2}-\frac{k_x^2(\epsilon_1+\epsilon_2)^2}{{\epsilon^{(0)}_{yz}}^2}\right),
\nonumber \\
\delta_{yz}&=&\frac{a_1^2 a_2^2 (\epsilon_1-\epsilon_2)^2 }
{12 (a_1+a_2)^2 \epsilon^{(0)}_{yz}} \frac{\omega^2}{c^2}. 
\label{eqDelt}
\end{eqnarray}

\noindent Note that since components of the wavevector are related to each other via Eq.(\ref{eqWave}), the choice of $k_x$ and $\omega/c$ as opposed to $k_z$ or $k_y$ [in Eq.(\ref{eqDelt})]  is somewhat arbitrary and primarily depends on the geometry. Here we use $k_x=2\pi j/h$ and $\omega$ as independent variables for $j^{\rm th}$ mode. 

The agreement between the non-local EMT with results of numerical solutions of Maxwell equations is shown in Fig.\ref{emt_nocorr}. It is clearly seen that Eq.(\ref{eqCorr}) perfectly describes the behavior of lower-order modes. The agreement tends to worsen for $n_{\rm eff}\gg 1$ where $|k \cdot a|\gtrsim 1$. To confirm that cladding regions and material absorption have weak effect on non-localities we have also simulated the modes of realistic Au-$\rm SiO_2$ structures with vacuum cladding regions. The results of these simulations and their comparison to our non-local EMT are shown in Fig.\ref{emt_nocorr}(c,d). 

An important note is that ``real'' parameter behind the validity of effective medium response is $|k \cdot a|\ll 1$. In majority of all-dielectric nanostructures $|\epsilon_{1,2}|\simeq 1$ or $|k_x c/\omega|\ll 1$, and this parameter is identical to the commonly-used criterion $a\ll \lambda$. For the high-index TM modes in metal-dielectric systems the spatial dispersion provides a significant correction to the quasi-static EMT results. The effective non-locality will be present for all-dielectric materials provided that $|\epsilon_{1,2}|\gg 1$. Similar effect have been recently discovered for microwave nanowire structures \cite{wires}. 

To conclude, we have demonstrated that conventional EMT fails to adequately describe the optical properties of multi-layered metal-dielectric metamaterials. We identified strong variation of the field to be the cause of this disagreement and derived an analytical correction to incorporate non-local effects into EMT. We have also demonstrated that multi-layered structures support high-index modes confined to spatial areas as small as $\lambda/8$. Our results, illustrated here for TM waves in two-component optical structures, can be used to design nano-guiding systems, and to provide an efficient link between the properties of realizable few-layer structures and their multi-layered macroscopic counterparts. The presented techniques are directly applicable to UV, IR, or THz metamaterials and can be generalized to mixed (HE,EH) waves and to multi-component systems involving anisotropic materials using techniques of \cite{Avrut2003}. 

This research was partially supported by PRF(ACS), GRF(OSU), and ONR


\end{document}